\documentclass[nohyperref]{article}

\usepackage{microtype}
\usepackage{graphicx}
\usepackage{booktabs} 

\usepackage{hyperref}

\usepackage{amsmath}
\usepackage{amssymb}
\usepackage{mathtools}
\usepackage{amsthm}

\usepackage{amsmath,amsfonts,bm}

\def\eqref#1{equation~\ref{#1}}

\def\1{\bm{1}}

\def\vv{{\bm{v}}}

\def\vx{{\bm{x}}}
\def\vy{{\bm{y}}}
\def\vz{{\bm{z}}}

\def\mX{{\bm{X}}}

\DeclareMathAlphabet{\mathsfit}{\encodingdefault}{\sfdefault}{m}{sl}
\SetMathAlphabet{\mathsfit}{bold}{\encodingdefault}{\sfdefault}{bx}{n}

\usepackage{subcaption}
\usepackage{caption}

\usepackage[accepted]{icml2022}

\usepackage[capitalize,noabbrev]{cleveref}

\theoremstyle{plain}

\theoremstyle{definition}

\theoremstyle{remark}

\usepackage[textsize=tiny]{todonotes}

\icmltitlerunning{Learning meaningful representations for radio astronomy with contrastive learning}

\begin{document}

\twocolumn[
\icmltitle{Learning useful representations for radio astronomy ``in the wild'' with contrastive learning}



\icmlsetsymbol{equal}{*}

\begin{icmlauthorlist}
\icmlauthor{Inigo Val Slijepcevic}{uom}
\icmlauthor{Anna M. M. Scaife}{uom,tur}
\icmlauthor{Mike Walmsley}{uom}
\icmlauthor{Micah Bowles}{uom}
\end{icmlauthorlist}

\icmlaffiliation{uom}{Department of Physics and Astronomy, University of Manchester, Manchester, UK}
\icmlaffiliation{tur}{Che Alan Turing Institute, 96 Euston Rd, London, UK}


\icmlcorrespondingauthor{Inigo Val Slijepcevic}{inigo.slijepcevic@postgrad.manchester.ac.uk}

\icmlkeywords{Machine Learning, ICML, radio astronomy, contrastive learning, self-supervised learning, representation learning}

\vskip 0.3in
]



\printAffiliationsAndNotice{\icmlEqualContribution} 

\begin{abstract}
Unknown class distributions in unlabelled astrophysical training data have previously been shown to detrimentally affect model performance due to dataset shift between training and validation sets. 
For radio galaxy classification, we demonstrate in this work that removing low angular extent sources from the unlabelled data before training produces qualitatively different training dynamics for a contrastive model. By applying the model on an unlabelled data-set with unknown class balance and sub-population distribution to generate a representation space of radio galaxies, we show that with an appropriate cut threshold we can find a representation with FRI/FRII class separation approaching that of a supervised baseline explicitly trained to separate radio galaxies into these two classes. Furthermore we show that an excessively conservative cut threshold blocks any increase in validation accuracy.  
We then use the learned representation for the downstream task of performing a similarity search on rare hybrid sources, finding that the contrastive model can reliably return semantically similar samples, with the added bonus of finding duplicates which remain after pre-processing.

\end{abstract}

\section{Introduction}
\label{sec:intro}

\begin{figure}
\centering
\begin{subfigure}[t]{.15\textwidth}
    \centering
    \includegraphics[width=0.95\linewidth]{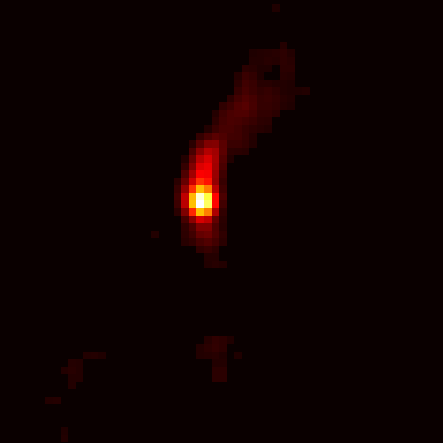}
    \caption{Fanaroff-Riley I}
\end{subfigure}%
\begin{subfigure}[t]{.15\textwidth}
    \centering
    \includegraphics[width=0.95\linewidth]{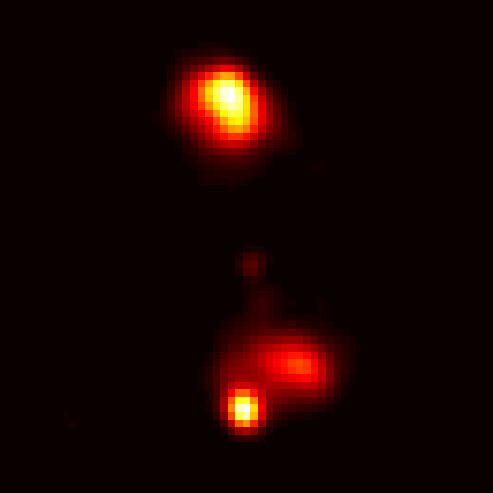}
    \caption{Fanaroff-Riley II}
\end{subfigure}
\caption{A typical example of each class cherry-picked from the MiraBest data-set. \cite{Miraghaei2017}.}
\label{figure:fr}
\end{figure}

To date in radio galaxy classification, supervised CNNs have largely been used to classify sources into the Fanaroff-Riley I (FRI) and FRII categories (Figure~\ref{figure:fr}), which has persisted as the canonical morphological distinction since it was established over 40 years ago \cite{Fanaroff1974}. However, such supervised classification requires many labelled samples. Acquiring labels is costly due to the labelling expertise required and may introduce selection biases training data are often chosen subject to observational factors such as brightness and distance constraints, \citealt{Hardcastle2020RadioJets}. Since our current understanding of the physics of radio galaxies is incomplete, there is also a question of whether the FR classification scheme is itself enforcing an implicit bias which may limit future scientific insight.

New radio sky surveys enabled by the construction of next-generation telescopes such as the Square Kilometre Array (SKA) will produce data on much larger scales than previous instruments, with significantly higher sensitivities expected to reveal a more morphologically diverse population of samples than are currently known. These data volumes will inevitably contain samples that challenge the current FR classification paradigm, and may be undetected or misclassified if our models remain tailored to the FR distinction. Furthermore, the volume of data, which is expected to be on the \textit{exabyte} scale, will likely be intractable even for citizen science projects such as Radio Galaxy Zoo \cite{Banfield2015} to label for supervised training. 

Semi-supervised learning has been explored as a potential solution to address these issues, as it leverages unlabelled data samples and allows for more data-driven separation in classification outputs \cite{Sohn2020FixMatch:Confidence, Berthelot2019MixMatch:Learning, Tarvainen2017MeanResults, Sellars2021LaplaceNet:Classification}. Semi-supervised learning without pre-training (consistency regularization \citet{Tarvainen2017MeanResults}; pseudo-labelling \citet{Pham2021Meta}, etc.) can make use of unlabelled data but is brittle to differences in the labelled and unlabelled data distributions, as has been shown both from a theoretical perspective \cite{vanEngelen2020ALearning} and empirically with astronomical data \cite{Slijepcevic2021CanClassification, Slijepcevic2022RadioShift}, making it less well-suited to training ``in the wild'' on real astronomical data.


Self-supervised representation learning has recently driven improvements in state-of-the-art performance on image classification tasks by leveraging large unlabelled data-sets for pre-training, with methods including SimCLR \cite{Chen2020ARepresentations}, MoCo \cite{HeMomentumLearning}, and BYOL \cite{Grill}; \citet{Jaiswal2020ALearning} provide a survey. Adapting pre-training for effective representation learning in the radio astronomy domain may allow us to use all of the available unlabelled data for training, rather than just a vanishingly small labelled fraction. Furthermore, it may help to provide flexibility for scientists to easily test hypotheses (such as alternate classification schema) with a significantly lower labelling cost by fine-tuning the pre-trained representation. Lastly, the use of techniques such as similarity search \cite{Stein2021Self-supervisedDatasets, Walmsley2022PracticalLearning} highlight the usefulness of a good representation space for applications beyond classification that require no labels at all.

\textbf{This work:} We focus on using contrastive learning without negative pairs (BYOL) to learn a semantically meaningful representation of radio galaxy images. We use the separation of the FRI/II classes as measured by the accuracy of a kNN classifier applied to the representation space to assess the quality of the representation, showing that our model separates the classes in a labelled validation data-set \textit{even when trained only on unlabelled data with an unknown distribution of labels and different selection cuts}. We compare the quality of this representation to the feature space learned by a supervised classifier with the same backbone architecture (Resnet18; \citet{He2016DeepRecognition}). We show that unresolved and low angular extent sources sources in the training data of the contrastive model negatively impact the learned representation, highlighting the importance of domain-specific data curation for self-supervised models, which in our case can be done automatically before training. We show that by feeding the encoder a single data-point we can extract similar data from the unlabelled data through a similarity search, highlighting the scientific usefulness of a semantically meaningful representation space beyond just classification.


\section{Data}
The Mirabest data-set contains 1256 radio galaxies, which have been selected for classification based on a variety of criteria \cite{Miraghaei2017}. The data-set is publicly available \cite{Porter2020MiraBestDataset} and has been widely used to train and evaluate machine learning models for radio galaxy classification. The images are classified into FRI/II classes and also contain \textit{Confident} and \textit{Uncertain} tags which are aligned with the confidence in the respective classification of the expert labellers. This is the dataset that we use for evaluating our model performance and to train the supervised baseline. The data-set also contains some \emph{hybrid} radio galaxy samples which we use only as input for a similarity search.

The Radio Galaxy Zoo Data Release 1 (RGZ DR1; Wong et al. in prep) contains $\sim 100,000$ sources  which have been obtained from the FIRST survey \cite{Becker1995TheCentimeters} for the RGZ citizen science project. Each image is paired with an arc-second extension (calculated algorithmically), which gives the projected angular size of the source in the sky. The RGZ DR1 data-set has an unknown FRI/II class balance and the abundance of sub-populations is also unknown. 
Since both the MiraBest and RGZ DR1 data-sets are drawn from the FIRST survey, we remove any shared samples from RGZ DR1.

\section{Method}
\label{sec:method}
Self-supervised learning attempts to learn a representation, $\vy_\theta$, given data, $\vx_i \in \mX$, by giving the model a task without labels but forces the model to learn a representation space which is useful for not-yet-known downstream tasks. The backbone of the model is an encoder, with the rest of model being discarded at inference: the set up is analogous to transfer learning without any initial labels. In our contrastive learning framework, the model is trained by being shown pairs of augmented images and rewarding augmentations of the same image being placed close together in the representation space of the encoder. Some algorithms, such as SimCLR \cite{Chen2020ARepresentations}, also teach the model to move augmentations of other images further away. However, recent work has shown that this is not required to achieve state-of-the-art accuracy, with Bootstrap Your Own Latent (BYOL) \cite{Grill} exceeding SimCLR classification performance on ImageNet. BYOL has significantly less computation time than algorithms with negative pairs and a much lower memory footprint, which allows us to iterate faster given our limited computation budget. Furthermore, our images have a fuzzier semantic meaning, with less clear cut differences between classes and a more fine-grained nature than computer vision data-sets, which may cause negative pairs to confound the model. For these reasons, we use BYOL as the core of our model.

BYOL uses a momentum encoder to calculate positive pair losses, which helps to avoid representation collapse, although it has been shown that this is not strictly required \cite{Chen2021ExploringLearning}. An exponential moving average of the online network parameters, $\theta$, gives the parameters of the momentum encoder, $\xi$, which are updated at each step such that

\begin{equation}
    \label{eq:momenc}
    \xi_i \leftarrow \tau \xi_{i-1} + (1-\tau) \theta,
\end{equation}

using Equation~\ref{eq:momenc}, where the hyperparameter $\tau$ gives the decay rate.

Both the momentum and online encoder have a fully connected projection head which output $\vz_\xi$ and $\vz_\theta$, respectively, with the online network also having an additional fully connected prediction head, $q_\theta$. These extra layers separate the loss from $\vy_\theta$, helping to learn a more generalisable representation space by preventing the model from overfitting to the loss function. This has been empirically shown to improve performance in the contrastive learning setting \cite{Chen2020ARepresentations}.

\begin{figure*}
\centering
    \centering
    \includegraphics[width=0.8\linewidth]{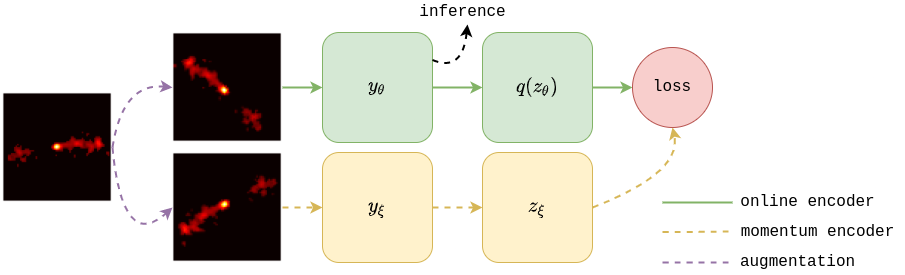}
    \caption{Diagram showing the flow of tensors in the BYOL algorithm. Dashed lines indicate that gradients are \textbf{not} back-propagated}
\label{figure:byol}
\end{figure*}

Two augmentations, $t$ and $t'$, are drawn from the same augmentation pool, $\mathcal{T}$, giving views $\vv$ and $\vv'$ of the original image, $\vx$. $\vv$ ($\vv'$) is passed through the online (momentum) encoder to give $q_\theta$ ($\vz_\theta$). A simple mean squared error between the momentum projection and online prediction of each augmentation gives the loss 
\begin{equation}
    \label{eq:byol_loss}
    \mathcal{L} = || q_\theta ( \vz_\theta ) - \vz_\eta ||^2_2,
\end{equation}
 which is back-propagated to update only the online parameters, $\theta$. The parameters of the momentum encoder, $\xi$,  are updated using the exponential moving average in Equation~\ref{eq:momenc}. Figure~\ref{figure:byol} illustrates the flow of tensors through the model.


\textbf{Training setup:} During training of BYOL, we only provide samples from the RGZ DR1 data-set, without showing the model any samples from MiraBest, and evaluate on \textit{Confident} samples from MiraBest. The supervised model is trained on MiraBest only, and evaluated on held out \textit{Confident} test samples. We use the representation learned by BYOL once the \textbf{training} loss has converged, rather than using early stopping to take the peak of the validation accuracy (which would likely give a higher test accuracy). We do this because we wish to emulate the setting where we are training a foundation model on unlabelled data, in which case there will be no labels with which to perform validation. During contrastive learning we modify the augmentations used in \citet{Chen2020ARepresentations} by reducing the size of the blurring kernel to 3 pixels and reducing the probability of blurring to 0.05, which empirically improves our results. We hypothesise that this is because of the sparseness of the radio galaxy images, where strong blurring results in distortion of semantically different fine-grained images and edges into similar looking blurs. We also reduce the range of sizes possible when cropping, to avoid cropping out important features of the image (e.g. jets). Both BYOL and the supervised baseline use the same ResNet-18 backbone with a fully connected layer at the end to downscale the feature space to a size of 100. BYOL uses multiple fully connected heads as explained above (details can be found in \citealt{Grill}), while the supervised model uses a simple fully connected classification head. \textbf{Hyperparameters:} cosine scheduler with SGD optimizer, learning rate: 0.6, momentum: 0.9, weight decay: 0.0005, batch size: 256.

\textbf{Evaluation:} we use a simple kNN classifier with 20 neighbours applied directly to the representation to evaluate the separation of the classes on a held out test set of \textit{Confident} MiraBest samples.

\textbf{Similarity search:} we perform a similarity search in the representation space by calculating the cosine similarity

\begin{equation}
    S^{j, input} = \frac{\vy_\theta^{input} \cdot \vy_\theta^{j}}{|\vy_\theta^{input}| | \vy_\theta^{j} | }
\end{equation}

of the input sample representation $\vy_\theta^{input}$ with the representation, $\vy_\theta^{j}$, of each data-point in the unlabelled data. The top $k$ images with the largest similarity scores are returned. 


\section{Results \& discussion}
\label{sec:Results}

\subsection{Model performance}

\begin{figure}
\centering
    \centering
    \includegraphics[width=0.8\linewidth]{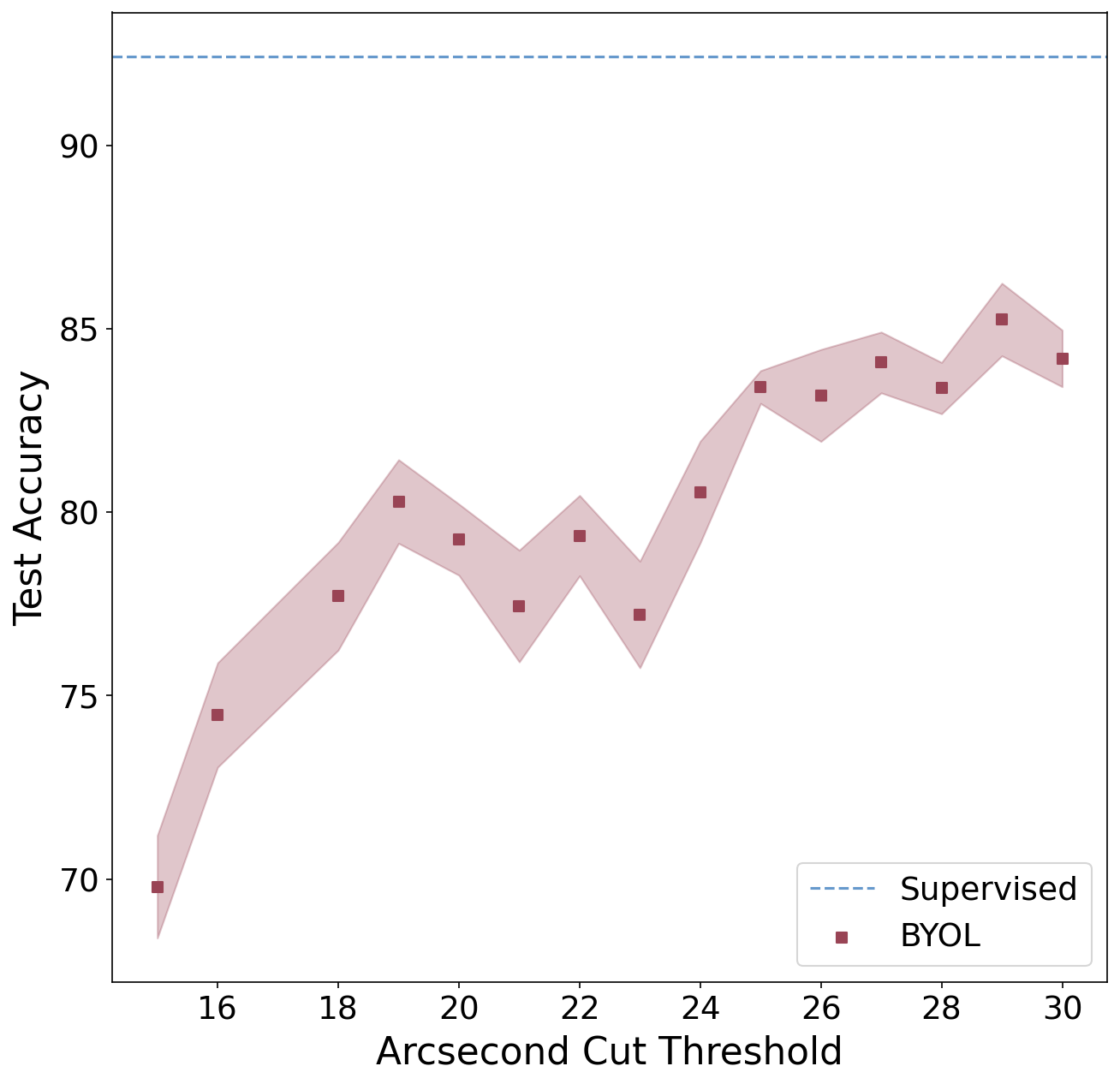}
    \caption{kNN accuracy as a function of cut threshold. Shaded area shows the standard error over 5 seeded runs.}
\label{figure:acc}
\end{figure}

\begin{figure}
\centering
    \centering
    \includegraphics[width=0.9\linewidth]{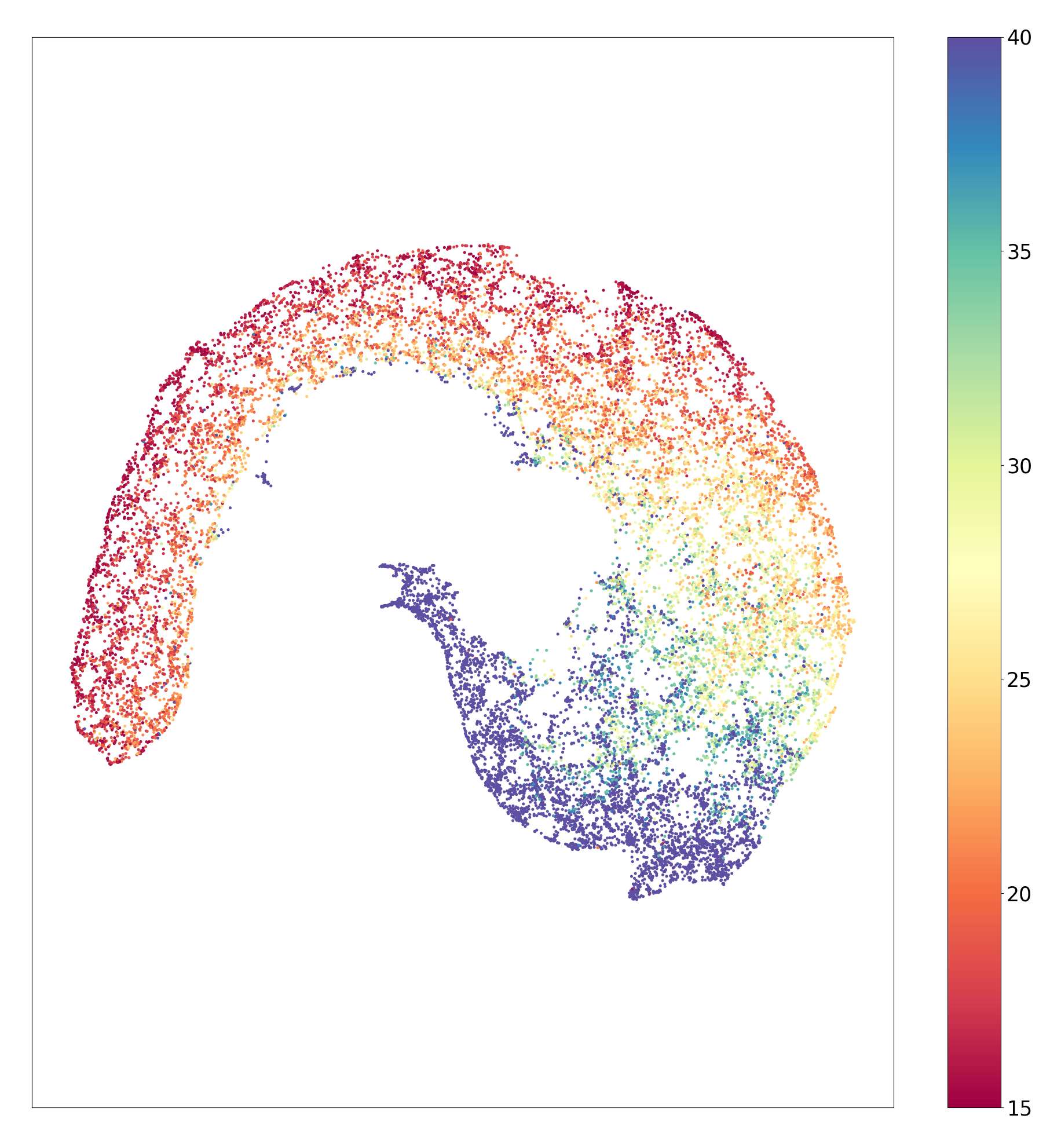}
    \caption{UMAP projection of the encoded representation RGZ DR1 data-set, with colorbar showing arcsecond extension.}
\label{figure:umap}
\end{figure}

We find that without seeing any MiraBest samples, BYOL learns a meaningful representation that separates out the \textit{Confident} MiraBest samples into FRI and FRII clusters. Figure~\ref{figure:acc} illustrates the representation quality compared with the supervised baseline, showing that the model is able to generalise from RGZ DR1 to MiraBest, which has been shown \textit{not} to be the case for a semi-supervised model using consistency regularization \cite{Slijepcevic2022RadioShift}. Although there is a gap between the supervised and self-supervised models (92.42\% vs. 85.25\%), it is surprisingly small given that the supervised model is \textit{explicitly trained to separate these classes on the same data-set being used in evaluation and with the benefit of labels}, whereas the self-supervised model has been trained on a different data-set with no labels. Furthermore, the representation learned by the contrastive model is likely to generalise better to a variety of downstream tasks due to the task-agnostic nature of the training. For example, in Figure~\ref{figure:umap} we see that our learned representation is structured with respect to source extension, which the model has no knowledge of during training. Lastly, any labels available post-hoc for a specific downstream classification task can still be used to fine-tune the network, which has been shown to be more data-efficient than training a supervised model from scratch \cite{Chen2020b}.

\begin{figure}
\centering
    \centering
    \includegraphics[width=0.9\linewidth]{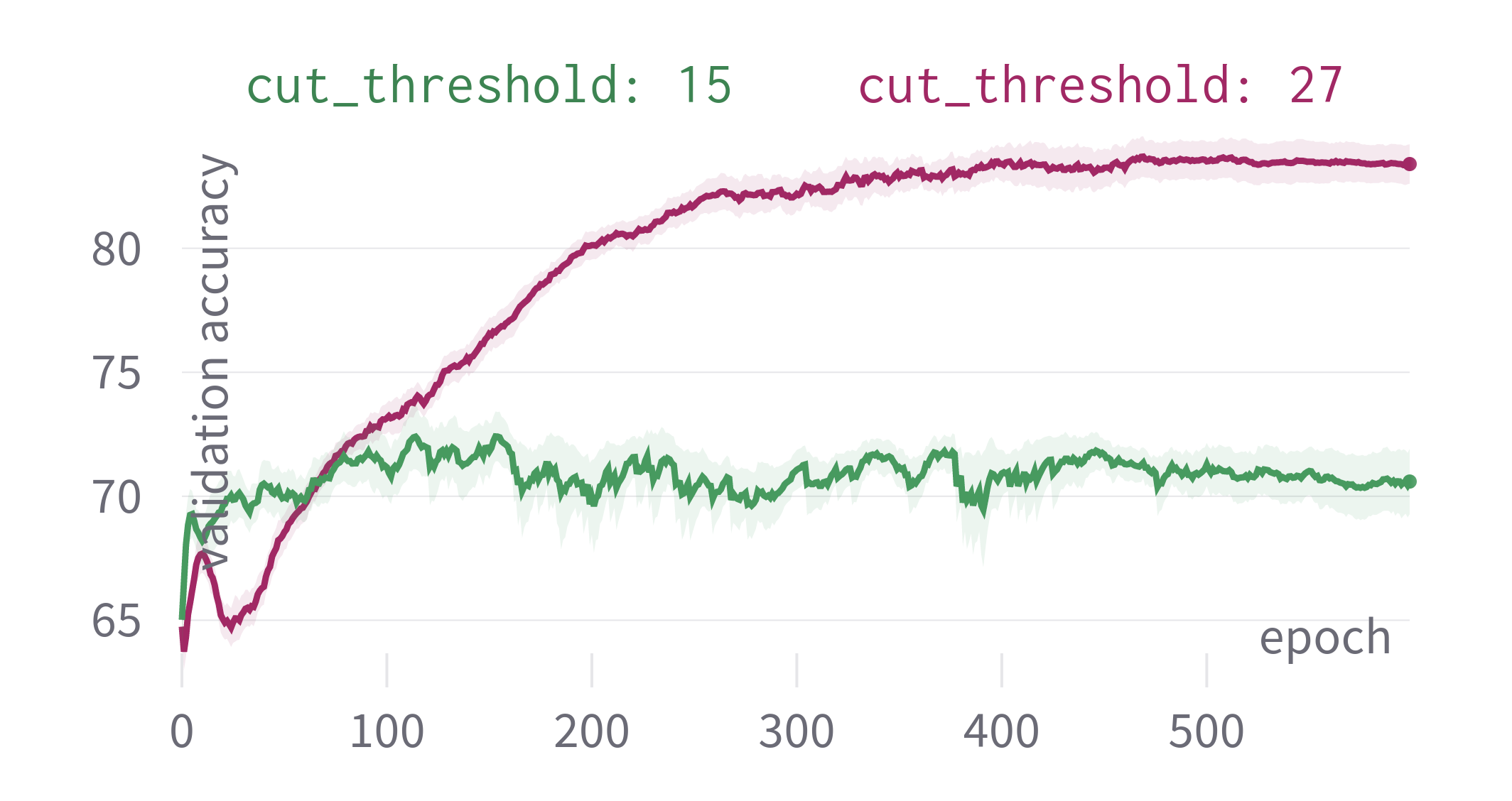}
    \caption{BYOL validation accuracy curves at different cut thresholds. Shaded area shows the standard error over 5 seeded runs, smoothed with an EMA constant of 0.7.}
\label{figure:cut_thresh}
\end{figure}

Figure~\ref{figure:acc} shows that choosing a suitable cut threshold is important, with the model's kNN classification performance suffering by over 10\% in the most extreme examples when low angular extent sources are included. Figure~\ref{figure:cut_thresh} shows that there is a qualitative difference in the way the model trains when the cut threshold is too low: at low cut thresholds the model is unable to improve its validation score even as the training loss decreases.

\subsection{Similarity search}

\begin{figure}
\centering
    \centering
    \includegraphics[width=0.66\linewidth]{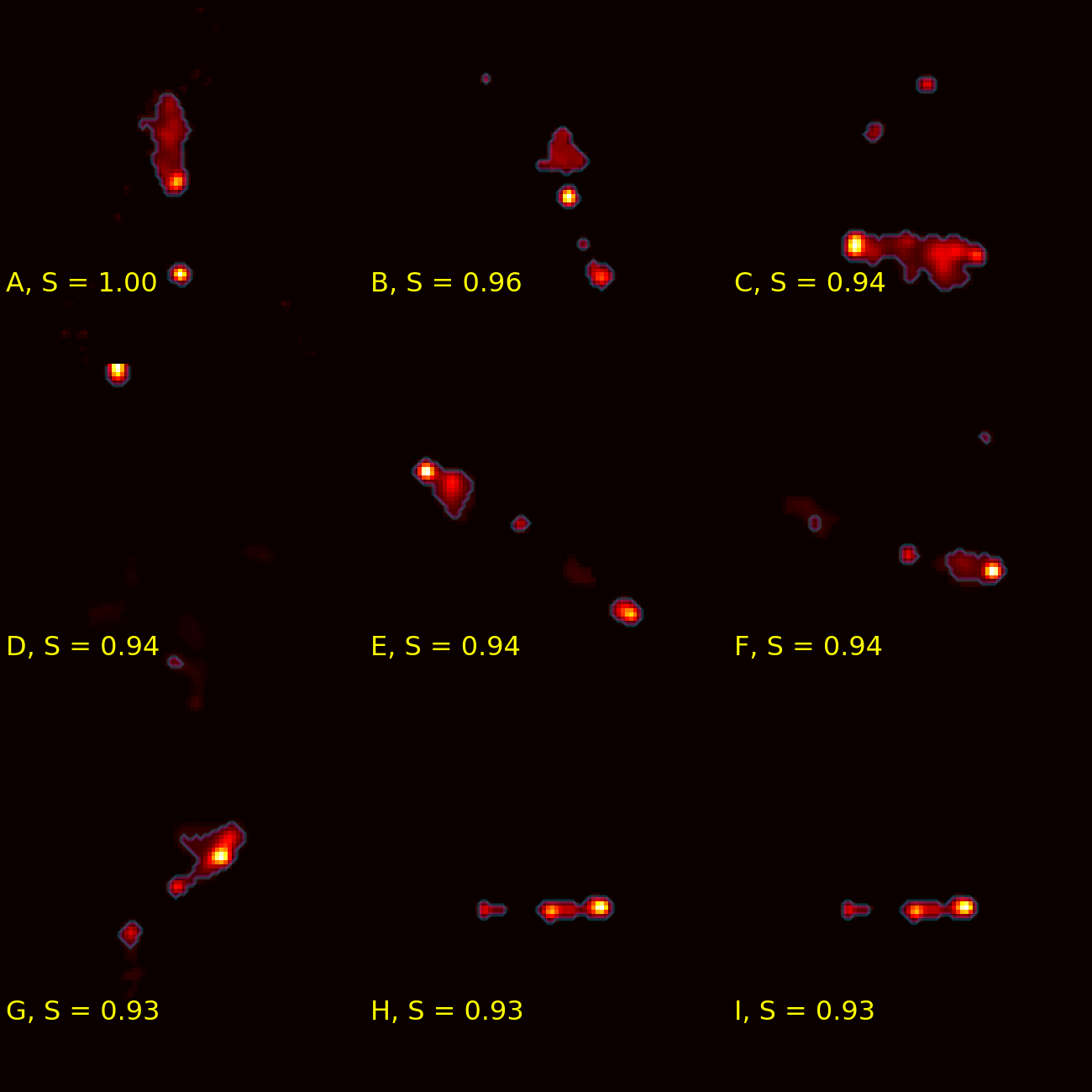}
    \caption{A similarity search performed on A (a rare hybrid galaxy from the MiraBest data-set) in the RGZ DR1 data-set. \textbf{Similar features:} (i) One extended diffuse jet. (ii) One bright spot from jet. (iii) Features extend to a similar angular scale. H and I are duplicates which evaded duplicate removal during pre-processing.
}
\label{figure:sim_search}
\end{figure}

In Figure~\ref{figure:sim_search} we feed our model with a hybrid image with both FRI and FRII properties from MiraBest, which would be considered out-of-distribution in the supervised case, and extract 3 nearest neighbours in the representation space. We see that we are able to extract semantically similar data-points from the RGZ DR1 data-set. We also find a duplicate data-point in the RGZ DR1 catalogue, see Figure~\ref{figure:sim_search}, even though the data have been (naively) pre-processed to remove duplicates. This shows that similarity search could also be used to correct errors where the same object has been included twice due to cataloging differences, e.g. slightly offset cutouts, which are difficult to catch in pre-processing.

\section{Conclusion}
We find that we are able to learn a semantically meaningful representation of radio galaxies with self-supervised learning. We we approach performance of the supervised case without any labels. This verifies that the representation learned is robust to data-set shift, indicating that we can use the vast quantities of available astronomical unlabelled data for representation learning \textit{without} the requirement of costly human classification and curation. We also find that some domain specific pre-processing is necessary for the radio galaxy case. While in our case this is easily applied, future work could investigate ways of automating this in a domain-agnostic way, which may also prove beneficial for training contrastive models in domains outside of astronomy.

We believe this work is a step towards developing large scale astronomical foundation models which can be shared and used by scientists for a variety of downstream tasks. Given the self-supervised nature of the learning task, the FRI/II separation and source extension structure in the representation space is indicative of a more general semantic structure of the vector space, and we expect that it may perform well on other physically meaningful tasks on similar data.

We show an example downstream task where we find images with similarly unusual physical properties to an out-of-distribution image by feeding it to the model and performing a similarity search in the representation space. This method could be used to find further examples of specific types of astronomical objects in a large unlabelled data-set and also to remove difficult-to-detect duplicates.

 \section{Acknowledgements}
We thank the anonymous reviewer whose comments improved this work.

IVS, AMS, MB \& MW gratefully acknowledge support from the UK Alan Turing Institute under grant reference EP/V030302/1. IVS gratefully acknowledges support from the Frankopan Foundation. HT gratefully acknowledges the support from the Shuimu Tsinghua Scholar Program of Tsinghua University. 

 This work has been made possible by the participation of more than 12,000 volunteers in the Radio Galaxy Zoo Project. The data in this paper are the result of the efforts of the Radio Galaxy Zoo volunteers, without whom none of this work would be possible. Their efforts are individually acknowledged at \url{ http://rgzauthors.galaxyzoo.org}.

\clearpage


\bibliography{references}

\begin{thebibliography}{23}
\providecommand{\natexlab}[1]{#1}
\providecommand{\url}[1]{\texttt{#1}}
\expandafter\ifx\csname urlstyle\endcsname\relax
  \providecommand{\doi}[1]{doi: #1}\else
  \providecommand{\doi}{doi: \begingroup \urlstyle{rm}\Url}\fi

\bibitem[Banfield et~al.(2015)Banfield, Wong, Willett, Norris, Rudnick,
  Shabala, Simmons, Snyder, Garon, Seymour, Middelberg, Andernach, Lintott,
  Jacob, Kapi{\'{n}}ska, Mao, Masters, Jarvis, Schawinski, Paget, Simpson,
  Kl{\"{o}}ckner, Bamford, Burchell, Chow, Cotter, Fortson, Heywood, Jones,
  Kaviraj, L{\'{o}}pez-S{\'{a}}nchez, Maksym, Polsterer, Borden, Hollow, and
  Whyte]{Banfield2015}
Banfield, J.~K., Wong, O.~I., Willett, K.~W., Norris, R.~P., Rudnick, L.,
  Shabala, S.~S., Simmons, B.~D., Snyder, C., Garon, A., Seymour, N.,
  Middelberg, E., Andernach, H., Lintott, C.~J., Jacob, K., Kapi{\'{n}}ska,
  A.~D., Mao, M.~Y., Masters, K.~L., Jarvis, M.~J., Schawinski, K., Paget, E.,
  Simpson, R., Kl{\"{o}}ckner, H.~R., Bamford, S., Burchell, T., Chow, K.~E.,
  Cotter, G., Fortson, L., Heywood, I., Jones, T.~W., Kaviraj, S.,
  L{\'{o}}pez-S{\'{a}}nchez, R., Maksym, W.~P., Polsterer, K., Borden, K.,
  Hollow, R.~P., and Whyte, L.
\newblock {Radio Galaxy Zoo: Host galaxies and radio morphologies derived from
  visual inspection}.
\newblock \emph{Monthly Notices of the Royal Astronomical Society},
  453\penalty0 (3):\penalty0 2326--2340, 2015.
\newblock ISSN 13652966.
\newblock \doi{10.1093/mnras/stv1688}.
\newblock URL \url{http://www.sepnet.ac.uk}.

\bibitem[Becker et~al.(1995)Becker, White, Helfand, Becker, White, and
  Helfand]{Becker1995TheCentimeters}
Becker, R.~H., White, R.~L., Helfand, D.~J., Becker, R.~H., White, R.~L., and
  Helfand, D.~J.
\newblock {The FIRST Survey: Faint Images of the Radio Sky at Twenty
  Centimeters}.
\newblock \emph{ApJ}, 450:\penalty0 559, 9 1995.
\newblock ISSN 0004-637X.
\newblock \doi{10.1086/176166}.
\newblock URL
  \url{https://ui.adsabs.harvard.edu/abs/1995ApJ...450..559B/abstract}.

\bibitem[Berthelot et~al.(2019)Berthelot, Carlini, Goodfellow, Oliver,
  Papernot, and Raffel]{Berthelot2019MixMatch:Learning}
Berthelot, D., Carlini, N., Goodfellow, I., Oliver, A., Papernot, N., and
  Raffel, C.
\newblock {MixMatch: A holistic approach to semi-supervised learning}.
\newblock Technical report, 2019.
\newblock URL \url{https://github.com/google-research/mixmatch}.

\bibitem[Chen et~al.(2020{\natexlab{a}})Chen, Kornblith, Norouzi, and
  Hinton]{Chen2020ARepresentations}
Chen, T., Kornblith, S., Norouzi, M., and Hinton, G.
\newblock {A Simple Framework for Contrastive Learning of Visual
  Representations}.
\newblock 2020{\natexlab{a}}.
\newblock URL \url{https://github.com/google-research/simclr.}

\bibitem[Chen et~al.(2020{\natexlab{b}})Chen, Kornblith, Swersky, Norouzi, and
  Hinton]{Chen2020b}
Chen, T., Kornblith, S., Swersky, K., Norouzi, M., and Hinton, G.
\newblock {Big self-supervised models are strong semi-supervised learners}.
\newblock \emph{Advances in Neural Information Processing Systems}, 2020-Decem,
  2020{\natexlab{b}}.
\newblock ISSN 10495258.
\newblock URL \url{https://github.com/google-research/simclr.}

\bibitem[Chen \& He(2021)Chen and He]{Chen2021ExploringLearning}
Chen, X. and He, K.
\newblock {Exploring Simple Siamese Representation Learning}.
\newblock pp.\  15745--15753, 2021.
\newblock \doi{10.1109/cvpr46437.2021.01549}.
\newblock URL \url{https://github.com/facebookresearch/simsiam}.

\bibitem[Fanaroff \& Riley(1974)Fanaroff and Riley]{Fanaroff1974}
Fanaroff, B.~L. and Riley, J.~M.
\newblock {The Morphology of Extragalactic Radio Sources of High and Low
  Luminosity}.
\newblock \emph{Monthly Notices of the Royal Astronomical Society}, 1974.
\newblock ISSN 0035-8711.
\newblock \doi{10.1093/mnras/167.1.31p}.

\bibitem[Grill et~al.(2020)Grill, Strub, Altch{\'{e}}, Tallec, Richemond,
  Buchatskaya, Doersch, Pires, Guo, Azar, Piot, Kavukcuoglu, Munos, and
  Valko]{Grill}
Grill, J.~B., Strub, F., Altch{\'{e}}, F., Tallec, C., Richemond, P.~H.,
  Buchatskaya, E., Doersch, C., Pires, B.~A., Guo, Z.~D., Azar, M.~G., Piot,
  B., Kavukcuoglu, K., Munos, R., and Valko, M.
\newblock {Bootstrap your own latent a new approach to self-supervised
  learning}.
\newblock In \emph{Advances in Neural Information Processing Systems}, volume
  2020-Decem, 2020.
\newblock ISBN 2006.07733v3.
\newblock URL
  \url{https://github.com/deepmind/deepmind-research/tree/master/byol}.

\bibitem[Hardcastle \& Croston(2020)Hardcastle and
  Croston]{Hardcastle2020RadioJets}
Hardcastle, M.~J. and Croston, J.~H.
\newblock {Radio galaxies and feedback from AGN jets}.
\newblock \emph{New Astronomy Reviews}, 88, 2020.
\newblock ISSN 13876473.
\newblock \doi{10.1016/j.newar.2020.101539}.
\newblock URL \url{http://www.jb.man.ac.uk/atlas/;}.

\bibitem[He et~al.()He, Fan, Wu, Xie, and Girshick]{HeMomentumLearning}
He, K., Fan, H., Wu, Y., Xie, S., and Girshick, R.
\newblock {Momentum Contrast for Unsupervised Visual Representation Learning}.
\newblock URL \url{https://github.com/facebookresearch/moco}.

\bibitem[He et~al.(2016)He, Zhang, Ren, and Sun]{He2016DeepRecognition}
He, K., Zhang, X., Ren, S., and Sun, J.
\newblock {Deep residual learning for image recognition}.
\newblock In \emph{Proceedings of the IEEE Computer Society Conference on
  Computer Vision and Pattern Recognition}, volume 2016-Decem, pp.\  770--778,
  2016.
\newblock ISBN 9781467388504.
\newblock \doi{10.1109/CVPR.2016.90}.
\newblock URL \url{http://image-net.org/challenges/LSVRC/2015/}.

\bibitem[Jaiswal et~al.(2020)Jaiswal, Babu, Zadeh, Banerjee, and
  Makedon]{Jaiswal2020ALearning}
Jaiswal, A., Babu, A.~R., Zadeh, M.~Z., Banerjee, D., and Makedon, F.
\newblock {A Survey on Contrastive Self-Supervised Learning}.
\newblock \emph{Technologies}, 9\penalty0 (1):\penalty0 2, 2020.
\newblock \doi{10.3390/technologies9010002}.

\bibitem[Miraghaei \& Best(2017)Miraghaei and Best]{Miraghaei2017}
Miraghaei, H. and Best, P.~N.
\newblock {The nuclear properties and extended morphologies of powerful radio
  galaxies: The roles of host galaxy and environment}.
\newblock \emph{Monthly Notices of the Royal Astronomical Society},
  466\penalty0 (4):\penalty0 4346--4363, 2017.
\newblock ISSN 13652966.
\newblock \doi{10.1093/mnras/stx007}.

\bibitem[Pham et~al.(2021)Pham, Dai, Xie, Luong, and Le]{Pham2021Meta}
Pham, H., Dai, Z., Xie, Q., Luong, M.-T., and Le, Q.~V.
\newblock {Meta Pseudo Labels}.
\newblock \emph{IEEE Conference on Computer Vision and Pattern Recognition},
  2021.
\newblock ISSN 2331-8422.
\newblock URL \url{http://arxiv.org/abs/2003.10580}.

\bibitem[Porter(2020)]{Porter2020MiraBestDataset}
Porter, F. A.~M.
\newblock {MiraBest Batched Dataset}.
\newblock \emph{https://zenodo.org/record/4288837}, 11 2020.
\newblock \doi{10.5281/ZENODO.4288837}.
\newblock URL \url{https://zenodo.org/record/4288837}.

\bibitem[Sellars et~al.(2021)Sellars, Aviles-Rivero, and
  Sch{\"{o}}nlieb]{Sellars2021LaplaceNet:Classification}
Sellars, P., Aviles-Rivero, A.~I., and Sch{\"{o}}nlieb, C.-B.
\newblock {LaplaceNet: A Hybrid Energy-Neural Model for Deep Semi-Supervised
  Classification}.
\newblock \emph{CoRR}, 2021.
\newblock URL \url{http://arxiv.org/abs/2106.04527}.

\bibitem[Slijepcevic \& Scaife(2021)Slijepcevic and
  Scaife]{Slijepcevic2021CanClassification}
Slijepcevic, I.~V. and Scaife, A. M.~M.
\newblock {Can semi-supervised learning reduce the amount of manual labelling
  required for effective radio galaxy morphology classification?}
\newblock \emph{NeurIPS 2021: Machine Learning and the Physical Sciences
  Workshop}, 11 2021.
\newblock URL \url{https://arxiv.org/abs/2111.04357v2
  http://arxiv.org/abs/2111.04357}.

\bibitem[Slijepcevic et~al.(2022)Slijepcevic, Scaife, Walmsley, Bowles, Wong,
  Shabala, and Tang]{Slijepcevic2022RadioShift}
Slijepcevic, I.~V., Scaife, A. M.~M., Walmsley, M., Bowles, M., Wong, I.,
  Shabala, S.~S., and Tang, H.
\newblock {Radio Galaxy Zoo: Using semi-supervised learning to leverage large
  unlabelled data-sets for radio galaxy classification under data-set shift}.
\newblock \emph{MNRAS}, \penalty0 (May), 2022.
\newblock URL \url{http://arxiv.org/abs/2204.08816}.

\bibitem[Sohn et~al.(2020)Sohn, Berthelot, Li, Zhang, Carlini, Cubuk, Kurakin,
  Zhang, and Raffel]{Sohn2020FixMatch:Confidence}
Sohn, K., Berthelot, D., Li, C.~L., Zhang, Z., Carlini, N., Cubuk, E.~D.,
  Kurakin, A., Zhang, H., and Raffel, C.
\newblock {FixMatch: Simplifying semi-supervised learning with consistency and
  confidence}.
\newblock In \emph{Advances in Neural Information Processing Systems}, volume
  2020-Decem, 2020.

\bibitem[Stein et~al.(2021)Stein, Harrington, Blaum, Medan, and
  Lukic]{Stein2021Self-supervisedDatasets}
Stein, G., Harrington, P., Blaum, J., Medan, T., and Lukic, Z.
\newblock {Self-supervised similarity search for large scientific datasets}.
\newblock 2021.
\newblock URL \url{http://arxiv.org/abs/2110.13151}.

\bibitem[Tarvainen \& Valpola(2017)Tarvainen and
  Valpola]{Tarvainen2017MeanResults}
Tarvainen, A. and Valpola, H.
\newblock {Mean teachers are better role models: Weight-averaged consistency
  targets improve semi-supervised deep learning results}.
\newblock \emph{Advances in Neural Information Processing Systems},
  2017-Decem:\penalty0 1196--1205, 2017.
\newblock ISSN 10495258.

\bibitem[van Engelen \& Hoos(2020)van Engelen and
  Hoos]{vanEngelen2020ALearning}
van Engelen, J.~E. and Hoos, H.~H.
\newblock {A survey on semi-supervised learning}.
\newblock \emph{Machine Learning}, 109\penalty0 (2), 2020.
\newblock ISSN 15730565.
\newblock \doi{10.1007/s10994-019-05855-6}.

\bibitem[Walmsley et~al.(2022)Walmsley, Scaife, Lintott, Lochner, Etsebeth,
  G{\'{e}}ron, Dickinson, Fortson, Kruk, Masters, Mantha, and
  Simmons]{Walmsley2022PracticalLearning}
Walmsley, M., Scaife, A. M.~M., Lintott, C., Lochner, M., Etsebeth, V.,
  G{\'{e}}ron, T., Dickinson, H., Fortson, L., Kruk, S., Masters, K.~L.,
  Mantha, K.~B., and Simmons, B.~D.
\newblock {Practical galaxy morphology tools from deep supervised
  representation learning}.
\newblock \emph{Monthly Notices of the Royal Astronomical Society}, 2022.
\newblock ISSN 0035-8711.
\newblock \doi{10.1093/mnras/stac525}.

\end{thebibliography}
\bibliographystyle{icml2022}

\end{document}